\documentclass[reprint,aps,prl,twocolumn ,groupedaddress,nobibnotes]{revtex4-1}
 \pdfoutput=1

\usepackage[english]{babel}
\usepackage{amsmath}
\usepackage{graphicx}
\usepackage{caption}
\usepackage{subcaption}
\usepackage[title]{appendix}
\usepackage{babel}
\usepackage{braket}

\begin{document}

	\title{Quantum Chaos in a Rydberg Atom System}
	\date{\today}
	\author{Yochai Werman}
	\affiliation{Department of Physics, University of California, Berkeley, CA 94720, USA}

	\begin{abstract}
A recent proposal by Hallam et al. suggested using the chaotic properties of the semiclassical equations of motion, obtained by the time dependent variational principle (TDVP), as a characterization of quantum chaos. In this paper, we calculate the Lyapunov spectrum of the semiclassical theory approximating the quantum dynamics of a strongly interacting Rydberg atom array, which lead to periodic motion. In addition, we calculate the effect of quantum fluctuations around this approximation, and obtain the escape rate from the periodic orbit. We compare this rate to the rate extracted from the exact solution of the quantum theory, and find an order of magnitude discrepancy. We conclude that in this case, chaos in the TDVP equations does not correpond to phsyical properties of the system. Our result complement those of Ho et al.\cite{Ho} regarding the escape rate from the semiclassical periodic orbit.
\end{abstract}

\maketitle

\section{Introduction}

The thermalization of closed quantum systems is one of the fundamental questions of quantum statistical physics\cite{Deutsch,Srednicki,Tasaki,Rigol}. Relatedly, the emergence of classical chaos from the underlying quantum mechanical laws is one of the least understood points in the quantum - classical correspondence. Thermalization of classical systems is linked to the chaotic nature of the classical equations of motion; the temporal evolution leads to an ergodic exploration of phase space, restricted only by conservation laws, and this exploration justifies the use of the microcanonical ensemble in the calculation of expectation values. Therefore, the thermalization rate of classical systems has to do with the quantitative features of chaos - Lyapunov exponents and the sum of the positive exponents, the Kolmogorov - Sinai (KS) entropy\cite{Kolmogorov,Sinai}.

Quantum systems, on the other hand, obey unitary dynamics, which are linear and non-chaotic; in addition, the notion of phase space does not exist. Rather, thermalization in quantum systems is often considered in the context of the Eigenstate Thermalization Hypothesis (ETH)\cite{Deutsch, Srednicki, D'Alessio}, according to which the thermal expectation values of local operators are encoded in the many-body eigenstates of the Hamiltonian, and the approach to thermal equilibrium is a process of dephasing between the different eigenstates whose superposition comprises the initial state. The fact that the microcanonical expectation value of a local operator is correctly predicted by a single (high energy) eigenstate means that the physical region which is probed by the local operator is thermalized with the rest of the system, which acts as a thermal bath; thus, the spatial entanglement of quantum many-body eigenstates plays a decisive role in quantum thermalization\cite{Polkovnikov1,Zurek,D'Alessio,Gogolin}.

There have been many theoretical attempts to find a bridge between the classical and quantum approaches to thermalization. An intriguing relation between classical chaos and quantum entanglement has been shown to hold in several single-body systems which have both a classical and a quantum realization: the rate of entanglement entropy growth after a quench in the quantum system\cite{Calabrese, Kim} has been shown to be related to the KS entropy in the classical system\cite{Zurek,Miller,Bianchi,Pattanayak,Monteoliva,Asplund,Bianchi1,Bianchi2}. 

Lately, Hallam et al.\cite{Hallam} have suggested a novel approach for understanding thermalization in many-body systems: using the time-dependent variational principle (TDVP)\cite{Haegeman}, it is possible to project the unitary quantum dynamics of a many-body system onto an effective semi-classical, Hamiltonian dynamics on a variational manifold. The emergent classical dynamics on such a manifolds exhibits chaotic behavior, and the Lyapunov exponents of this semiclassical system may be numerically compared to properties of the full quantum system. In particular, Hallam et al. have found a connection between the KS entropy of the semiclassical variational approximation and the entanglement entropy growth rate of the quantum system in the transverse-field Ising model.

\begin{figure}[htp]
        \center{\includegraphics
        {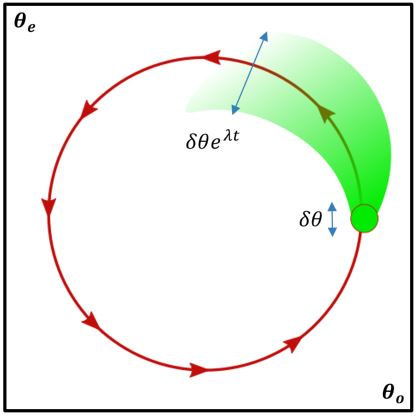}}
        \caption{\label{fig:decayrate} Calculation of the decay rate of a periodic orbit in phase space. The decay rate depends on the Lyapunov exponents of the flow, and on the uncertainty in the initial conditions, dictated by quantum fluctuations. We consider a generalized phase space, consisting of variational parameters, and Hamiltonian equations of motion obtained from the quantum theory using TDVP.}
      \end{figure}

In this work, we extend this line of reasoning for an experimentally relevant system: strongly interacting Rydberg atoms arranged on a one-dimensional lattice\cite{NatureLukin}. Such a system has been theoretically analyzed, and using a clever variational matrix product state (MPS) description, Bernien et al.\cite{NatureLukin} have obtained a Hamiltonian which offers a semi-classical description of the many-body Rydberg array. Utilizing this emergent Hamiltonian, we calculate the decay rate of a particular initial state in a manner depicted in Fig.~\ref{fig:decayrate}; we obtain the magnitude of quantum fluctuations in the initial conditions using the truncated Wigner approximation (TWA)\cite{Polkovnikov2}, and multiply it by the KS entropy which we obtain by computing the Lyapunov spectrum. We find that the decay rate predicted using this semi-classical approximation is an order of magnitude smaller than the exact rate, calculated numerically; we conclude that for this system, the TDVP does not capture the main mechanism responsible for the decay rate, or thermalization.

\section{Model}

In a recent experiment, Bernien et al.\cite{NatureLukin} have managed to construct a flawless one - dimensional array of cold $^{86}Rb$ ions. These atoms are subjected to an electromagnetic field which, for a single atom, facilitates Rabi oscillations between the ground state $|g\rangle$ and excited Rydberg state $|r\rangle$ of the atom. If the spacing between the atoms is small enough, the repulsive van der Waals interactions between neighboring atoms give rise to the phenomenon of Rydberg blockade - the simultaneous excitation of nearest neighbors is suppressed. The Hamiltonian of the system under consideration (known colloquially as the $PXP$ model) is given by

\begin{eqnarray}
H = \frac{1}{2}\Omega\sum_i P_{i-1}^g \sigma^x_i P_{i+1}^g,
\end{eqnarray}

where $\sigma^x_i = |g_i\rangle\langle r_i| + |r_i\rangle\langle g_i|$ creates transitions between ground and Rydberg state on site $i$, with Rabi frequency $\Omega$; the projectors $P^g_i = |g_i\rangle\langle g_i|$ are the consequence of the Rydberg blockade.

Experiment\cite{NatureLukin} and numerical simulations\cite{Turner, Turner2, Ho, Choi} on small systems revealed that the relaxation under unitary dynamics specified by this Hamiltonian strongly depends on the initial state of the system. Starting from a generic initial state, the system shows fast relaxation and no revivals, characteristic of thermalizing systems. However, starting from the $Z_2$ state, which consists of alternating ground- and Rydberg- state ions, $|Z_2\rangle = |g_1 r_2 g_3...\rangle$, the system shows surprising long-time oscillations of local observables. This seemingly non-ergodic behavior of a high-energy initial state has been termed ``quantum many body scar''\cite{Turner,Turner2}, and has to do with a small (order $N$, where $N$ is the number of atoms) subset of non-ergodic eigenstates. This surprising behavior of the $Z_2$ initial state, which displays long lived oscillations and very slow equilibration, is the focus of our work.

The theoretical description of the Rydberg atom system is facilitated by a sophisticated representation in terms of a matrix product state. Previous work\cite{NatureLukin} has suggested a representation of this system using a bond dimension $\chi = 2$ MPS:

\begin{widetext}
\begin{eqnarray}
|\psi(\{\theta\})\rangle = \sum_{\{\sigma_i\}}v_L^\dagger A(\theta_1,\phi_1)^{\sigma_1}A(\theta_2,\phi_2)^{\sigma_2}...A(\theta_L,\phi_L)^{\sigma_L}v_R|\{\sigma_i\}\rangle,
\end{eqnarray}
\end{widetext}

where $\sigma_i = g_i / r_i$ refers to the ground or Rydberg state, the corresponding matrices are

\begin{eqnarray}
A(\theta,\phi)^g = \begin{pmatrix}
    \cos(\theta)       & 0 \\
   1       & 0 \\
\end{pmatrix}, A(\theta,\phi)^r = \begin{pmatrix}
   0       & ie^{i\phi}\sin(\theta) \\
   0      & 0 \\
\end{pmatrix}
\end{eqnarray}

and $v_L^\dagger = (1, 0)$, $v_R^\dagger = (1, 1)$. The matrices are parameterized by angles $\theta$ and $\phi$, where $\cos(\theta)$ is the amplitude for the ground state configuration, $\sin(\theta)$ the amplitude for the Rydberg state, and $\phi$ the relative phase. This MPS representation automatically satisfies the Rydberg blockade constraint, as $A^r\cdot A^r = 0$.
 
TDVP yields semiclassical equations of motion for the variational variables $\theta_i$ and $\phi_i$. For the initial state $|Z_2\rangle$, translational invariance dictates that only two sets of angles $, \phi_1$ and $\theta_2, \phi_2$ for the even and odd sites, should be considered. The classical initial conditions for such a state are $\theta_1 = \pi, \phi_2 = \phi_1 = 0$; the initial value of $\theta_2$ is not well defined for the $Z_2$ state. These angles obey the semiclassical equations of motion (see Appendix for derivation)

\begin{eqnarray}
\dot{\theta}_1 &=& -\frac{1}{2}\Omega\left[\sin(\theta_1)\cos^2(\theta_1)\tan(\theta_2) + \cos(\theta_2)\right]\nonumber\\
\dot{\theta}_2 &=& -\frac{1}{2}\Omega\left[\sin(\theta_2)\cos^2(\theta_2)\tan(\theta_1) + \cos(\theta_1)\right]\nonumber\\
\phi_1&=& \phi_0 \text{ }= \text{ }0.
\end{eqnarray}

A numerical solution of these variational equations for the $Z_2$ initial state results in a periodic motion with a frequency of approximately $\Omega/1.51$, with the manybody wavefunction oscillating between two staggered configurations; this solution is in surprisingly good agreement with experiment and exact numerical solution of the quantum dynamical equations; thus, for the $Z_2$ state, the semiclassical approximation is reliable, at least for short times, of the order of $\Omega^{-1}$.

The semiclassical equations obtained by the $\chi = 2$ TDVP predict a periodic orbit that does not decay. The experiment, and more exact solutions of the quantum dynamics, display a decay of the oscillations in the Rydberg density, up to a long time behavior which does not display such oscillations. In this work, we calculate the decay rate predicted due to the lowest order quantum fluctuations around the semiclassical expansion; this is done using the truncated Wigner approximation (TWA). To this end, we calculate the Lyapunov spectrum of the classical periodic orbit, and the Wigner function of the initial $|Z_2\rangle$ state. We find that the decay rate due to quantum fluctuations is much smaller than that observed in exact numerics; we conclude that the semiclassical approximation does not offer an accurate description of this process, and in particular the KS entropy of the semiclassical flow is not related to the entanglement entropy growth rate of the quantum system.

\section{Truncated Wigner Approximation}

According to the TWA, quantum fluctuations enter only through the initial conditions, while the equations of motion are affected only in higher orders in $\hbar$. In particular, the expectation value of an observable operator $\hat{O}$ at time $t$, given an initial density matrix $\rho_0$ at time $t_0$, will be given by

\begin{widetext}
\begin{eqnarray}
\langle \hat{O}(t)\rangle = \int D\xi(t_0)\int D\xi(t) W(\xi(t_0)) O_W(\xi(t)) \delta\left[\xi(t)-\xi_{cl}(t,\xi(t_0))\right] + O(\hbar^2),
\end{eqnarray}
\end{widetext}

where in this expression $D\xi(t_0), D\xi(t)$ stand for integration over the initial and final semiclassical variables, $W(\xi)$ is the Wigner function of the initial density matrix, $O_W(\xi)$ is the Weyl symbol of the operator $\hat{O}$, and $\xi_{cl}(t,\xi(t_0))$ is the solution of the semiclassical equations of motion at time $t$, given the initial conditions $\xi(t_0)$. An intuitive interpretation of this result is that, in the leading order in quantum fluctuations, the expectation value of an operator is the average of the Weyl symbol of the operator at time $t$ weighted with the Wigner function, which acts as a probability distribution for the initial conditions; however, the Wigner function may be negative, and thus is not a true probability distribution.

The Wigner function for the initial density matrix and the Weyl symbol for the operator are the semiclassical phase space descriptions for the respective quantum operators. The quantum uncertainty principle between canonically conjugate variables manifests itself as a minimum width of the functions in phase space. Thus, the larger the quantum fluctuations, the broader will be the Wigner function. In addition, negative values of the Wigner function are clear signatures of quantum mechanics.

The Wigner function arises naturally from the path integral formulation of quantum dynamics. As explained in detail in the Appendix, the Wigner function of the $Z_2$ initial state is given by (there is no dependence on the $\phi$ degrees of freedom for this initial state)

\begin{widetext}
\begin{eqnarray}
W(\theta_1,\theta_2)& =& \frac{1}{4}K(\theta_1,\theta_2)[1-\sqrt{3}\cos(\vartheta(\theta_1,\theta_2))][1+\sqrt{3}\cos(\vartheta(\theta_2,\theta_1))]
\end{eqnarray}

where

\begin{eqnarray}
\vartheta(\theta_1,\theta_2) &=& 2\tan^{-1}\left[\frac{\tan\left(\frac{\theta_1}{2}\right)}{\cos\left(\frac{\theta_2}{2}\right)}\right]\\
K(\theta_1,\theta_2) &=& \frac{1 - \sin^2\left(\frac{\theta_1}{2}\right)\sin^2\left(\frac{\theta_2}{2}\right)}{\left[\cos^2\left(\frac{\theta_1}{2}\right) + \tan^2\left(\frac{\theta_2}{2}\right)\right]\left[\cos^2\left(\frac{\theta_2}{2}\right) + \tan^2\left(\frac{\theta_1}{2}\right)\right]\cos\left(\frac{\theta_1}{2}\right)\cos\left(\frac{\theta_2}{2}\right)}\nonumber\\
&\times&\frac{\sin(\vartheta_1(\theta_1,\theta_2))\sin(\vartheta_2(\theta_1,\theta_2))}{\sin(\theta_1)\sin(\theta_2)}\nonumber
\end{eqnarray}
\end{widetext}

\begin{figure}[htp]
\centering
\begin{subfigure}{.55\textwidth}
  \centering
  \includegraphics[width=0.8\linewidth]{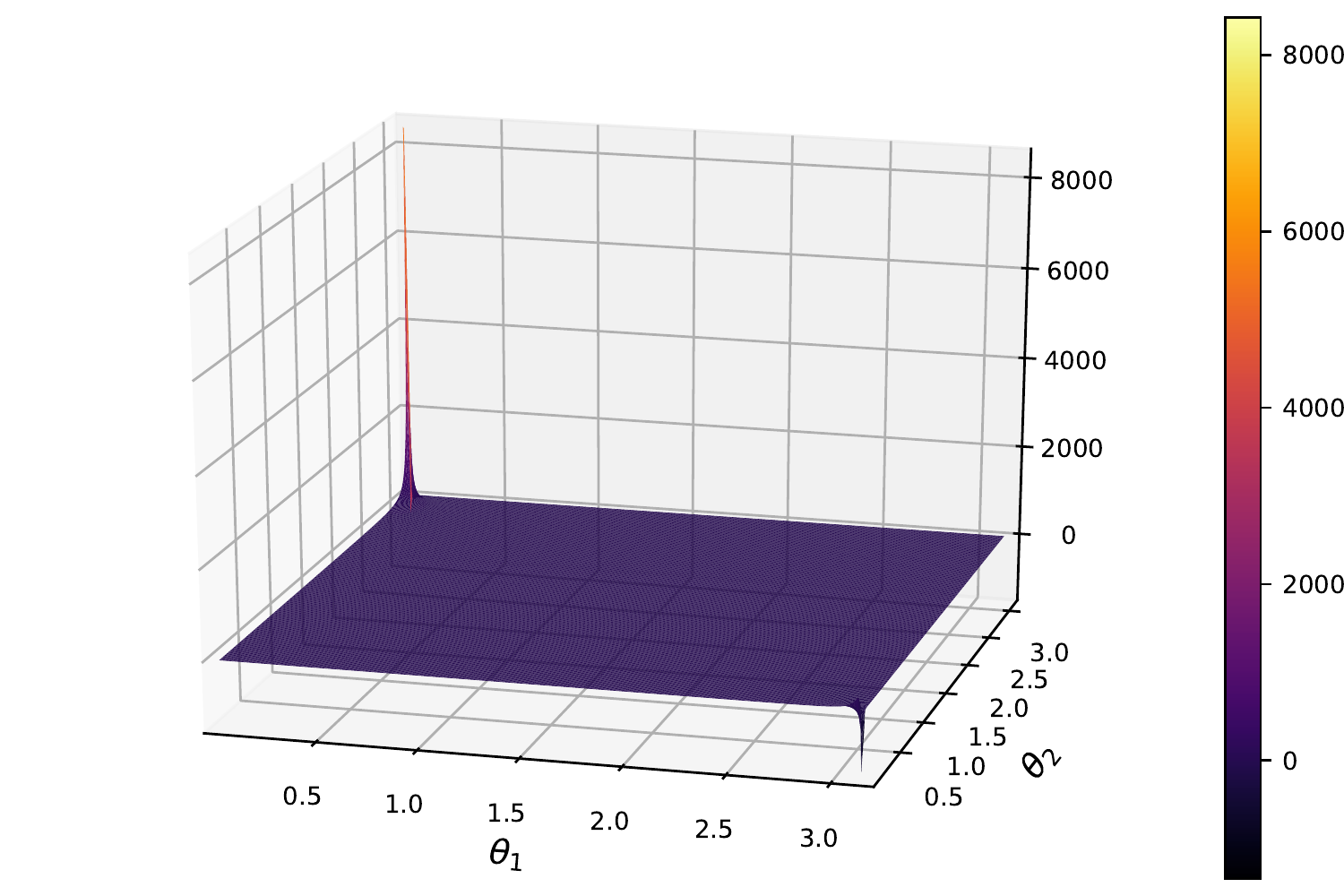}
  \caption{Wigner function in constrained Hilbert space. }
  \label{fig:sub1}
\end{subfigure}
\begin{subfigure}{.55\textwidth}
  \centering
  \includegraphics[width=0.8\linewidth]{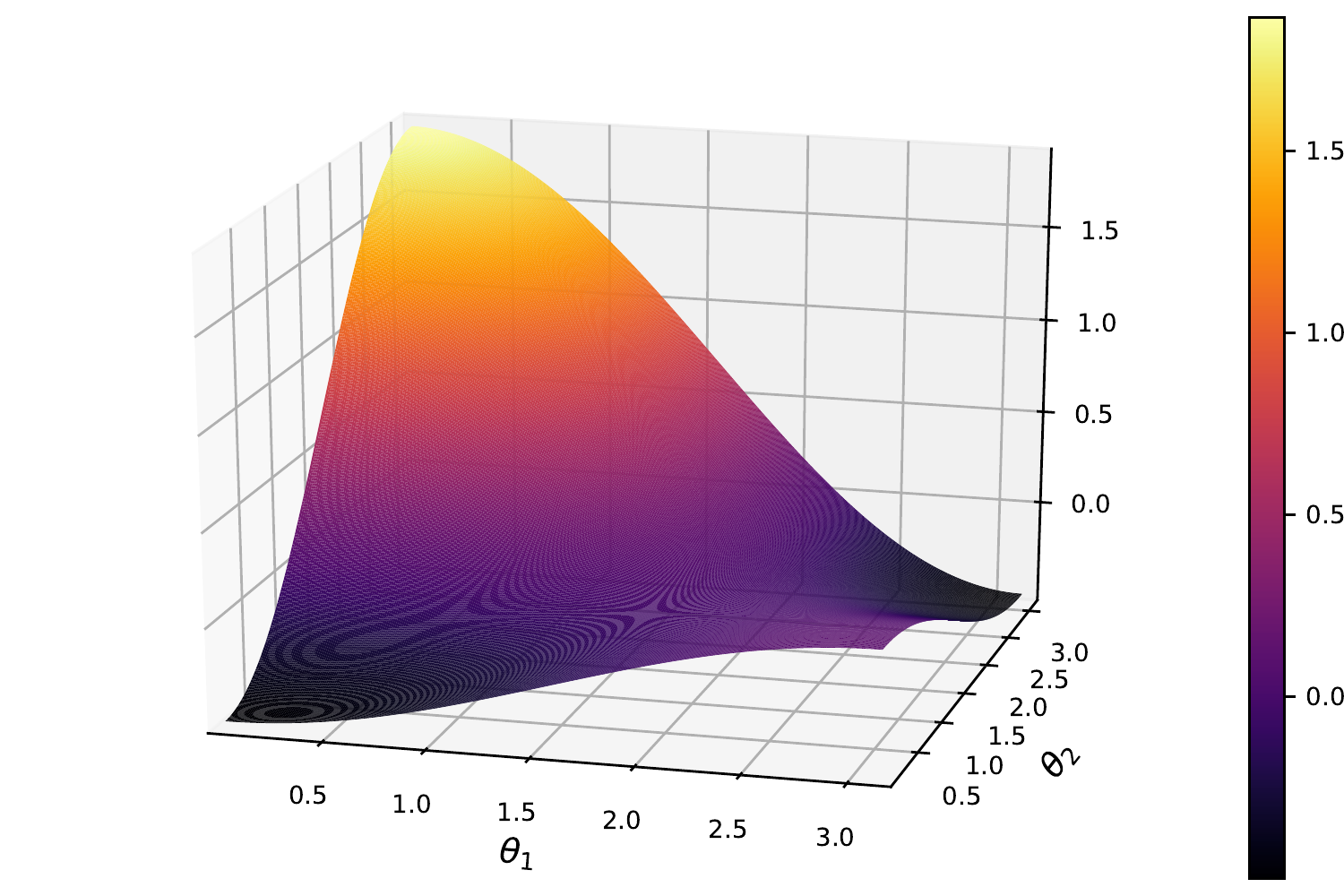}
  \caption{Wigner function in an unconstrained Hilbert space.}
  \label{fig:sub2}
\end{subfigure}
\caption{Wigner function for the initial $|Z_2\rangle = |g\rangle_1|r\rangle_2$ state. In the constrained Hilbert space, the Wigner function is strongly peaked near the classical initial values $\theta_1 = 0, \theta_2 = \pi$. In comparison, for the same initial state in an unconstrained Hilbert space, the Wigner function is very broad, and is non-negligible throughout the entire phase space. Thus, the constrained dynamics suppress quantum fluctuations in the evolution of the $|Z_2\rangle$ state, rendering the semi-classical approximation accurate for long times.}
\label{fig:Wigner}
\end{figure}

The complicated form of the Wigner function arises from the constraint on nearest neighbor excitations; this form of the Wigner function may be compared to the Wigner function for the state $|\tilde{Z}_2\rangle = \ket{\downarrow_1\uparrow_2}$ in an unconstrained Hilbert space:

\begin{eqnarray}
\tilde{W}(\theta_1,\theta_2)& =& \frac{1}{4}[1-\sqrt{3}\cos(\theta_1)][1+\sqrt{3}\cos(\theta_2)]
\end{eqnarray}

A plot of the Wigner function is shown in Fig.~\ref{fig:Wigner}, and compared to the Wigner function of the $\tilde{Z}_2$ state in an unconstrained Hilbert space. The Wigner function in the constrained Hilbert space is very sharply peaked about the classical value $\theta_1 = 0, \theta_2 = \pi$, in contrast to the Wigner function of the $\tilde{Z}_2$ state in an unconstrained Hilbert space, which is broad due to the large quantum fluctuations of a spin-$1/2$ system. The fact that the Wigner function is strongly peaked suggests that the TWA is a good approximation to the exact quantum dynamics. This uncertainty in the initial conditions may be considered as an effective $\hbar$ - it is a small parameter which arises due to the quantum nature of the system.

\section{Lyapunov spectrum}
The trajectory which starts at the phase point $\theta_1=\phi_1=\phi_2=0, \theta_2=\pi$ is known; it desribes a periodic motion associated with the periodic oscillations of the quantum Rydberg system. We are now interested in examining the behavior of nearby trajectories, with initial values that differ slightly from this value, and remain within the small width of the Wigner function. The Lyapunov exponents of a flow in phase space are a quantitative measure of the exponential separation of adjacent trajectories. Consider a phase point, denoted by $\mathbf{X}$, which is subject to Hamiltonian equations of motion, which may be summarized as

\begin{eqnarray}
\dot{\mathbf{X}}(t) = \mathbf{F}(\mathbf{X}(t)).
\end{eqnarray}

The solution of the equations of motion then determine the trajectory of the phase point given some initial value $\mathbf{X}(0)$. The deviation of a trajectory initially separated from $\mathbf{X}(0)$ by $\delta\mathbf{X}(0)$ is given to linear order in the deviation by

\begin{eqnarray}
\delta\dot{\mathbf{X}}(0) = \frac{\partial\mathbf{F}}{\partial\mathbf{X}}\cdot\delta\mathbf{X}(t);
\end{eqnarray}

this equation is formally solved by

\begin{eqnarray}
\delta\mathbf{X}(t) &=& \mathbf{M}(t,\mathbf{X}(0))\cdot\delta\mathbf{X}(0),\nonumber\\
\mathbf{M}(t,\mathbf{X}(0) ) &=& T\exp\left[\int_0^t\frac{\partial\mathbf{F}}{\partial\mathbf{X}}\right],
\end{eqnarray}
where $T$ denotes time ordering.

The Lyapunov exponents, which depend on the initial point, are defined as the logarithm of the eigenvalues of the matrix $\mathbf{M}(t,\mathbf{X}(0))$, in the limit of infinite time; if this matrix has eigenvalues $\{\chi_i(t,\mathbf{X}(0))\}$, the exponents are defined as

\begin{eqnarray}
\lambda_i(\mathbf{X}(0)) = \lim_{t\rightarrow\infty}\frac{1}{t}\log\chi_i(t,\mathbf{X}(0)).
\end{eqnarray}

This generalizes intuitively for periodic orbits of period $T$, on which

\begin{eqnarray}
\lambda_i(\mathbf{X}(0)) = \frac{1}{T}\log\chi_i(T,\mathbf{X}(0)).
\end{eqnarray}

Positive Lyapunov exponents are associated with the expanding directions in phase-space, negative ones with contracting directions, and zero Lyapunov exponents with neutral directions. The direction in phase-space along the Hamiltonian flow is a neutral direction, as it does not expand exponentially. For Hamiltonian equations of motion, the Lyapunov exponents come in pairs with opposite signs. 

We have calculated the Lyapunov spectrum for this flow, to perturbations with differing wave-vectors. The exponents are shown in Fig.~\ref{fig:exponents}; details of the calculation are shown in the appendix. Surprisingly, we find that the dominant Lyapunov exponent corresponds to the $Z_2$ invariant perturbation. As expected, we find that the Lyapunov exponents are much smaller than the energy scale of the Hamiltonian. If this were not the case, we would not have expected to observe such a striking footprint of the semiclassical Hamiltonian dynamics on the exact quantum dynamics; in addition, this motivates the term "quantum many-body scars". 

\begin{figure}[htp]
        \center{\includegraphics[width=0.5\textwidth]
        {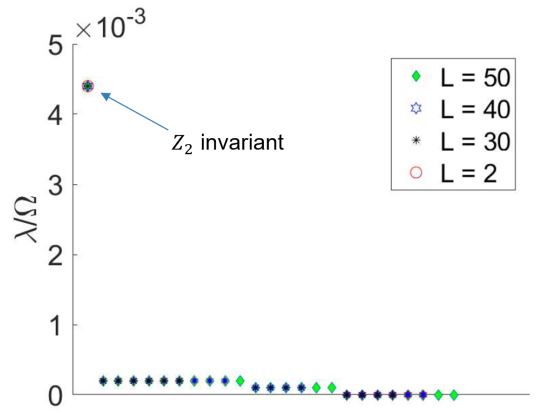}}
        \caption{\label{fig:exponents} The Lyapunov exponents for varying sizes of the unit cell of the perturbation. The Lyapunov exponents are small compared to the energy scale of the Hamiltonian; this is to be expected, as the periodic orbit in the semiclassical approximation has a striking signature in the quantum mechanical evolution of the $|Z_2\rangle$ state. Surprisingly, the largest exponent appears for the $Z_2$ invariant excitation, and appears already in the smallest unit cell.}
      \end{figure}

The Kolmogorov - Sinai entropy measures the exponential rate in which dynamical systems produce information. According to Pesin's theorem, it is equal to the sum of positive Lyapunov exponents:

\begin{eqnarray}
h_{KS} = \sum_i\Theta(\lambda_i)\lambda_i.
\end{eqnarray}

$h_{KS}$ measures the rate in which an initial infinitesimal volume in phase space is stretched in the direction of the positive Lyapunov exponents. We calculate the KS entropy for the semiclassical equations of motion, and for a system of $30$ atoms we get

\begin{eqnarray}
h_{KS} = 0.006.
\end{eqnarray}

\section{Escape rate}
Having calculated the Wigner function and the KS entropy, we are now set to calculate the escape rate from the periodic orbit. As a deviation grows exponentially in chaotic classical dynamics, schematically

\begin{eqnarray}
\delta\theta(t) = \delta\theta_0e^{\sum_i\lambda_i t} =\delta\theta_0e^{h_{KS} t} ,
\end{eqnarray}

the time at which the deviation becomes significant is 

\begin{eqnarray}
t^* = \frac{1}{h_{KS}}\log\delta\theta_0^{-1},
\end{eqnarray}

corresponding to an escape rate

\begin{eqnarray}
\Lambda =  \frac{1}{\log[\delta\theta_0^{-1}]}h_{KS}.
\end{eqnarray}

In our analysis, $\delta\theta_0$ is given by the width of the Wigner function, which gives the uncertainty in the initial conditions; as an estimate, we set $\delta\theta_0\sim 0.01$. Using the results of the previous sections, we conclude that $\Lambda \approx 0.001 $ for a system consisting of thirty Rydbern atoms.

\begin{figure}[htp]
\centering
\begin{subfigure}{.5\textwidth}
  \centering
  \includegraphics[width=0.8\linewidth]{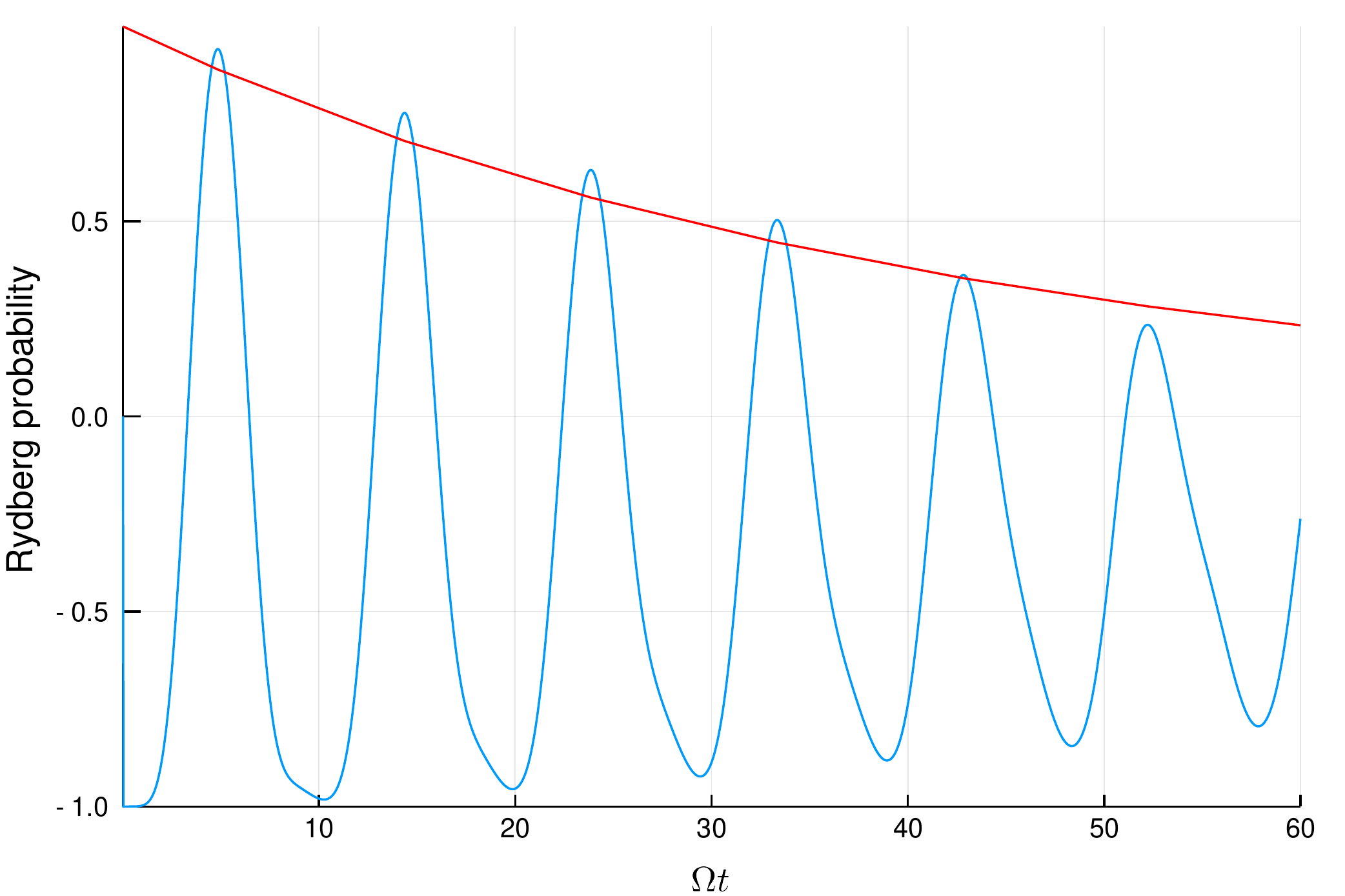}
  \caption{Decay in the revivals of the Rydberg probability on site $N/2$. }
  \label{fig:sub1}
\end{subfigure}
\begin{subfigure}{.5\textwidth}
  \centering
  \includegraphics[width=0.8\linewidth]{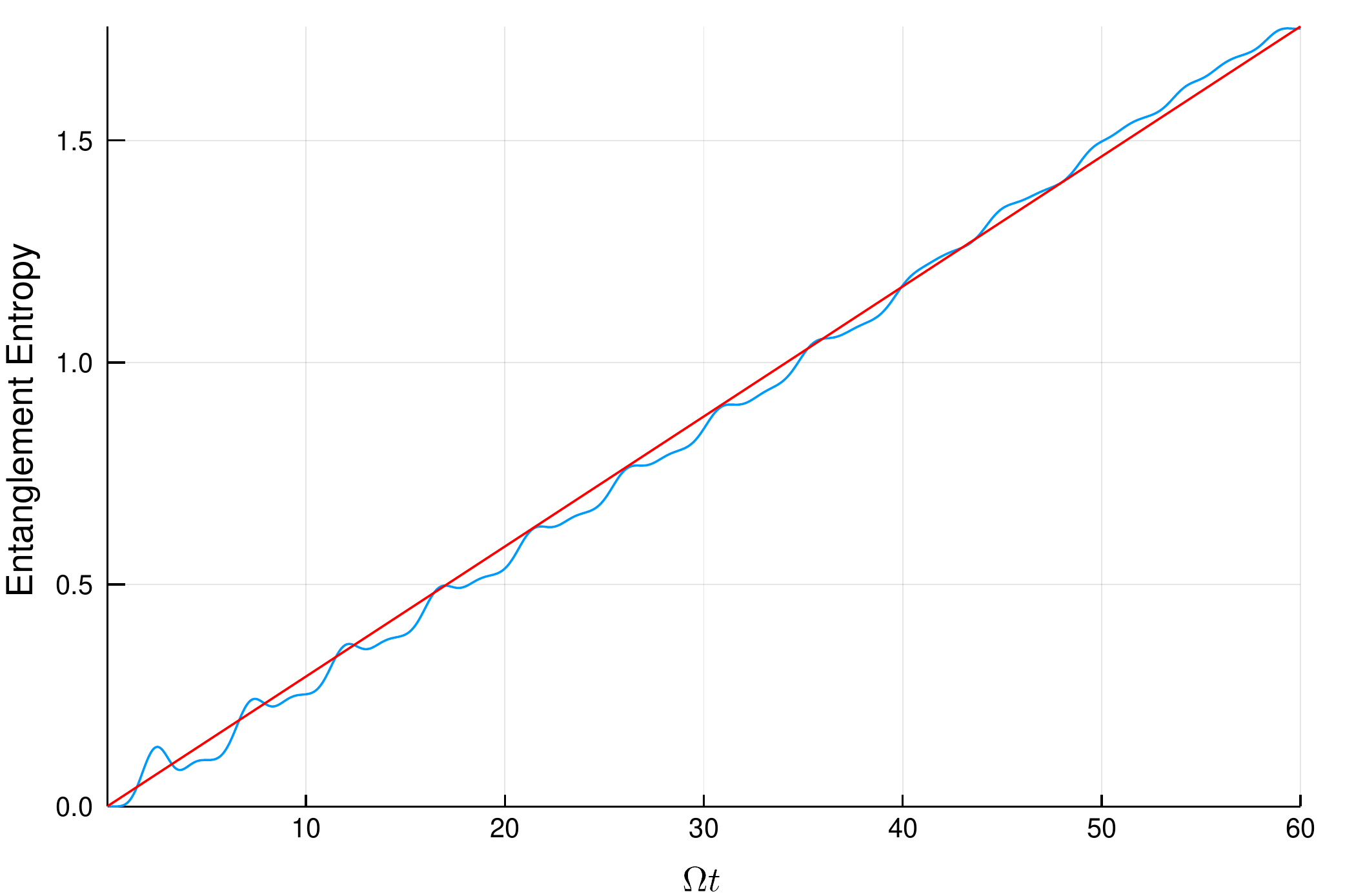}
  \caption{Growth of entanglement entropy for half the system.}
  \label{fig:sub2}
\end{subfigure}
\begin{subfigure}{.5\textwidth}
  \centering
  \includegraphics[width=0.8\linewidth]{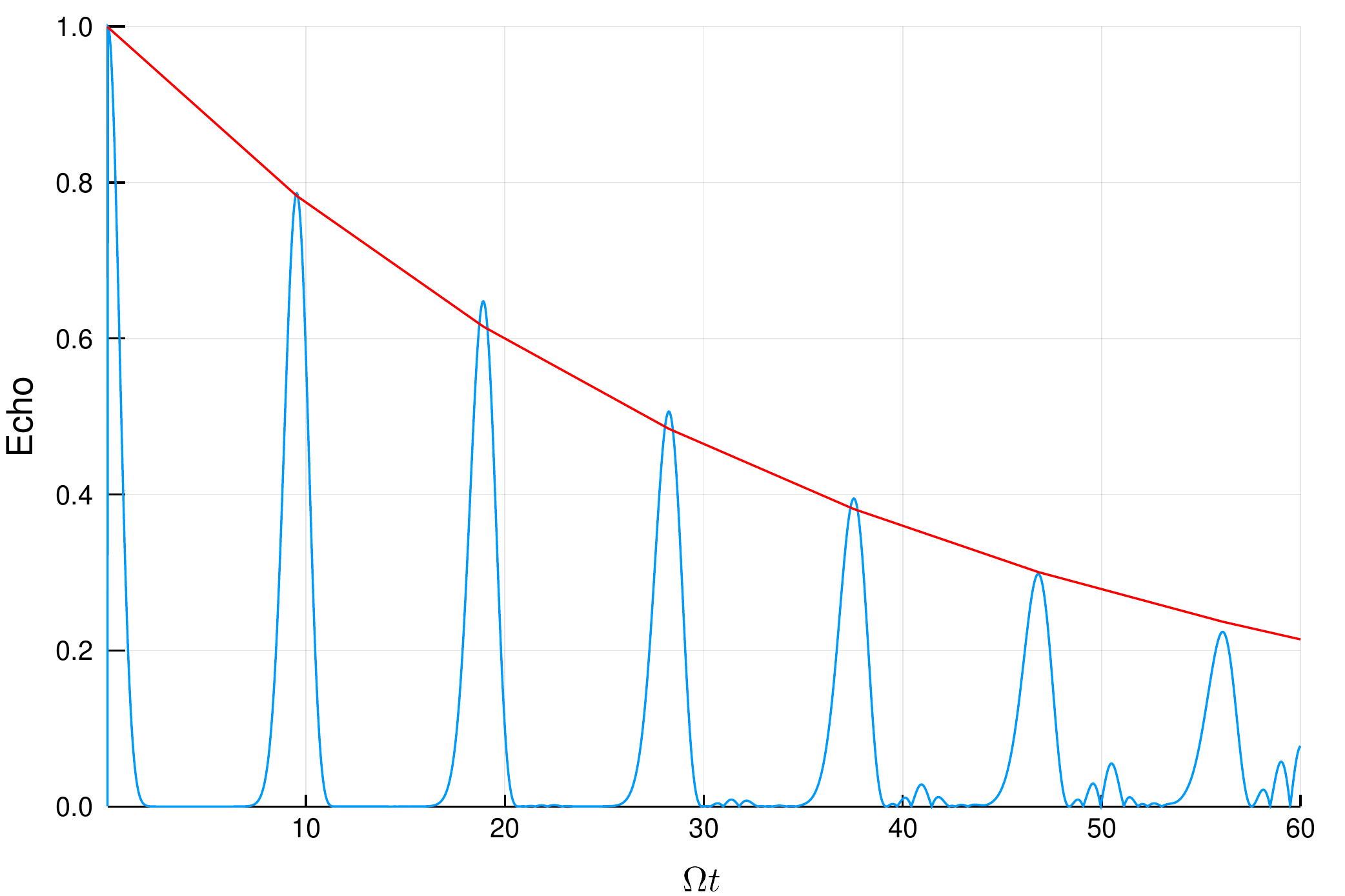}
  \caption{Decay in the Lochschmidt Echo.}
  \label{fig:sub2}
  \end{subfigure}
\caption{The fact that the dynamics of the Rydberg system is not exactly periodic can be observed from the decay in the oscillations of the on-site Rydberg probability, from the growth of the entanglement entropy, or from the decay of the Lochschmidt echo $|\langle\psi(t)|\psi(0)\rangle|$. The blue lines correspond to the solution of the quantum dynamics using TEBD, while the red lines are exponential (for the echo and decay of oscillations) or linear (for the entanglement entropy) fits.}
\label{fig:quantumsolution}
\end{figure}

We compare the escape rate, which we calculated within the semiclassical theory using a bond dimension $\chi = 2$ variational approximation, to the exact escape rate, obtained form a numerical solution of the quantum dynamics. The exact escape rate can be extracted from the decay rate of the periodic revivals in Rydberg density, from the rate of increase in entanglement entropy, or from the decay of the Lochschmidt echo, $|\langle\psi(t)|\psi(0)\rangle|$. We used the time-evolving block decimation (TEBD) technique to solve for the dynamics for a system of thirty atoms, and calculate the three quantities, which are shown in Fig~\ref{fig:quantumsolution}. The extracted escape rate from the three quantities (corresponding to the exponential decay rate of the oscillations in the Rydberg probability and Lochschmidt echo, and to the linear increase in the entanglement entropy) is about $0.02-0.03$, orders of magnitude larger than that predicted by the semiclassical approximation.

\section{Discussion}

In this work, we have analyzed the well-known $PXP$ model, relevant to recent experiments with Rydeberg atoms. Bernien et al.\cite{NatureLukin} have used the TDVP to obtain a semiclassical Hamiltonian, which approximates the exact quantum unitary dynamics. Using this semiclassical description, we calculated the Lyapunov spectrum related to the periodic evolution of the $Z_2$ state, for perturbations of various wave-vectors. In addition, we computed the magnitude of quantum fluctuations to lowest order in $\hbar$, through the width of the Wigner function related to the initial $Z_2$ state. Using these results, we arrived at the decay rate of the periodic solution.

We found that the decay rate calculated in the semiclassical approximation is orders of magnitude smaller than that obtained from the exact quantum dynamics; this is surprising, as the TDVP is particularly apt in obtaining the period of oscillations observed in the time evolution. We conclude that although the variational manifold captures the periodic nature of the oscillations, the decay of these oscillations involves motion in directions perpendicular to this manifold.

This conclusion leads us to claim that even if some aspects of the quantum unitary dynamics are well captured by TDVP, to some order of approximation, this does not imply that all the pertinent information appears to the same order. In particular, we find that the entanglement entropy growth for the $Z_2$ initial state is orders of magnitude larger than the KS entropy, in contrast to previous results. Thus, the applicability of a semiclassical description in the elucidation of processes of decoherence and thermalization must be further elucidated.

Ho et al.\cite{Ho} have analyzed the escape rate from the variational manifold along the orbit, defined as

\begin{eqnarray}
\epsilon_C = \oint_C \gamma(\theta_2(t),\theta_1(t))dt, \mbox{with}\nonumber\\
\gamma(\mathbf{\theta}) = ||\left(iH+\mathbf{\dot{\theta}}\partial_\mathbf{\theta}\right)|\psi(\mathbf{\theta})\rangle||;
\end{eqnarray}

they found that $\epsilon_C\approx 0.17$, which is smaller than the escape rate found on other trajectories corresponding to different initial states, yet much larger than the KS entropy we find in our calculation. This further establishes the fact that although the short time quantum dynamics are well captured within the $\xi=2$ variational manifold, the escape from the periodic orbit, with its concomitant growth in the entanglement entropy, is due to motion perpendicular to this manifold, and is not related to the Lyapunov spectrum of the motion. 

\bibliography{paper}

\begin{thebibliography}{10}

\bibitem{Asplund}
Curtis~T. Asplund and David Berenstein.
\newblock Entanglement entropy converges to classical entropy around periodic
  orbits.
\newblock {\em Annals of Physics}, 366:113 -- 132, 2016.

\bibitem{NatureLukin}
Hannes Bernien, Sylvain Schwartz, Alexander Keesling, Harry Levine, Ahmed
  Omran, Hannes Pichler, Soonwon Choi, Alexander~S. Zibrov, Manuel Endres,
  Markus Greiner, Vladan Vuletić, and Mikhail~D. Lukin.
\newblock Probing many-body dynamics on a 51-atom quantum simulator.
\newblock {\em Nature}, 551:579, 11 2017.

\bibitem{Bianchi1}
Eugenio Bianchi, Lucas Hackl, and Nelson Yokomizo.
\newblock Entanglement entropy of squeezed vacua on a lattice.
\newblock {\em Phys. Rev. D}, 92:085045, Oct 2015.

\bibitem{Bianchi}
Eugenio Bianchi, Lucas Hackl, and Nelson Yokomizo.
\newblock Linear growth of the entanglement entropy and the kolmogorov-sinai
  rate.
\newblock {\em Journal of High Energy Physics}, 2018(3):25, Mar 2018.

\bibitem{Bianchi2}
Eugenio Bianchi, Lucas Hackl, and Nelson Yokomizo.
\newblock Linear growth of the entanglement entropy and the kolmogorov-sinai
  rate.
\newblock {\em Journal of High Energy Physics}, 2018(3):25, Mar 2018.

\bibitem{Calabrese}
Pasquale Calabrese and John Cardy.
\newblock Evolution of entanglement entropy in one-dimensional systems.
\newblock {\em Journal of Statistical Mechanics: Theory and Experiment},
  2005(04):P04010, apr 2005.

\bibitem{Choi}
Soonwon Choi, Christopher~J. Turner, Hannes Pichler, Wen~Wei Ho, Alexios~A.
  Michailidis, Zlatko Papi\ifmmode~\acute{c}\else \'{c}\fi{}, Maksym Serbyn,
  Mikhail~D. Lukin, and Dmitry~A. Abanin.
\newblock Emergent su(2) dynamics and perfect quantum many-body scars.
\newblock {\em Phys. Rev. Lett.}, 122:220603, Jun 2019.

\bibitem{D'Alessio}
Luca D'Alessio, Yariv Kafri, Anatoli Polkovnikov, and Marcos Rigol.
\newblock From quantum chaos and eigenstate thermalization to statistical
  mechanics and thermodynamics.
\newblock {\em Advances in Physics}, 65(3):239--362, 2016.

\bibitem{Deutsch}
J.~M. Deutsch.
\newblock Quantum statistical mechanics in a closed system.
\newblock {\em Phys. Rev. A}, 43:2046--2049, Feb 1991.

\bibitem{Gogolin}
Christian Gogolin and Jens Eisert.
\newblock Equilibration, thermalisation, and the emergence of statistical
  mechanics in closed quantum systems.
\newblock {\em Reports on Progress in Physics}, 79(5):056001, apr 2016.

\bibitem{Haegeman}
Jutho Haegeman, J.~Ignacio Cirac, Tobias~J. Osborne, Iztok
  Pi\ifmmode~\check{z}\else \v{z}\fi{}orn, Henri Verschelde, and Frank
  Verstraete.
\newblock Time-dependent variational principle for quantum lattices.
\newblock {\em Phys. Rev. Lett.}, 107:070601, Aug 2011.

\bibitem{Hallam}
A.~Hallam, J.~G. Morley, and A.~G. Green.
\newblock The lyapunov spectra of quantum thermalisation.
\newblock {\em Nature Communications}, 98:2708, Jun 2019.

\bibitem{Ho}
Wen~Wei Ho, Soonwon Choi, Hannes Pichler, and Mikhail~D. Lukin.
\newblock Periodic orbits, entanglement, and quantum many-body scars in
  constrained models: Matrix product state approach.
\newblock {\em Phys. Rev. Lett.}, 122:040603, Jan 2019.

\bibitem{Kim}
Hyungwon Kim and David~A. Huse.
\newblock Ballistic spreading of entanglement in a diffusive nonintegrable
  system.
\newblock {\em Phys. Rev. Lett.}, 111:127205, Sep 2013.

\bibitem{Kolmogorov}
A.~N. Kolmogorov.
\newblock A new metric invariant of transient dynamical systems and
  automorphisms in lebesgue spaces.
\newblock {\em Dokl. Akad. Nauk SSSR}, 119:861--864, 1958.

\bibitem{Miller}
Paul~A. Miller and Sarben Sarkar.
\newblock Signatures of chaos in the entanglement of two coupled quantum kicked
  tops.
\newblock {\em Phys. Rev. E}, 60:1542--1550, Aug 1999.

\bibitem{Monteoliva}
Diana Monteoliva and Juan~Pablo Paz.
\newblock Decoherence and the rate of entropy production in chaotic quantum
  systems.
\newblock {\em Phys. Rev. Lett.}, 85:3373--3376, Oct 2000.

\bibitem{Pattanayak}
Arjendu~K. Pattanayak.
\newblock Lyapunov exponents, entropy production, and decoherence.
\newblock {\em Phys. Rev. Lett.}, 83:4526--4529, Nov 1999.

\bibitem{Polkovnikov2}
Anatoli Polkovnikov.
\newblock Phase space representation of quantum dynamics.
\newblock {\em Annals of Physics}, 325(8):1790 -- 1852, 2010.

\bibitem{Polkovnikov1}
Anatoli Polkovnikov, Krishnendu Sengupta, Alessandro Silva, and Mukund
  Vengalattore.
\newblock Colloquium: Nonequilibrium dynamics of closed interacting quantum
  systems.
\newblock {\em Rev. Mod. Phys.}, 83:863--883, Aug 2011.

\bibitem{Rigol}
Marcos Rigol, Vanja Dunjko, and Maxim Olshanii.
\newblock Thermalization and its mechanism for generic isolated quantum
  systems.
\newblock {\em Nature}, 452:854--858, Apr 2008.

\bibitem{Sinai}
Y.~Sinai.
\newblock {K}olmogorov-{S}inai entropy.
\newblock {\em Scholarpedia}, 4(3):2034, 2009.
\newblock revision \#91407.

\bibitem{Srednicki}
Mark Srednicki.
\newblock Chaos and quantum thermalization.
\newblock {\em Phys. Rev. E}, 50:888--901, Aug 1994.

\bibitem{Tasaki}
Hal Tasaki.
\newblock From quantum dynamics to the canonical distribution: General picture
  and a rigorous example.
\newblock {\em Phys. Rev. Lett.}, 80:1373--1376, Feb 1998.

\bibitem{Turner2}
C.~J. Turner, A.~A. Michailidis, D.~A. Abanin, M.~Serbyn, and
  Z.~Papi\ifmmode~\acute{c}\else \'{c}\fi{}.
\newblock Quantum scarred eigenstates in a rydberg atom chain: Entanglement,
  breakdown of thermalization, and stability to perturbations.
\newblock {\em Phys. Rev. B}, 98:155134, Oct 2018.

\bibitem{Turner}
C.~J. Turner, A.~A. Michailidis, D.~A. Abanin, M.~Serbyn, and Z.~Papić.
\newblock Weak ergodicity breaking from quantum many-body scars.
\newblock {\em Nature Physics}, 14:745--749, Jul 2018.

\bibitem{Zurek}
Wojciech~Hubert Zurek and Juan~Pablo Paz.
\newblock Decoherence, chaos, and the second law.
\newblock {\em Phys. Rev. Lett.}, 72:2508--2511, Apr 1994.

\end{thebibliography}

\newpage
\widetext
\appendix

\section{Appendix}
\section{Wigner function}
\subsection{Wigner function for a single spin coherent state}
Following Polkovnikov\cite{Polkovnikov2}, we first derive the well-known form of the Wigner function for a spin-$1/2$ coherent state

\begin{eqnarray}
W(\theta,\phi) =Tr\left[\hat{\rho}\hat{U}(\theta,\phi)\hat{\Pi}\hat{U}(\theta,\phi)^\dagger\right],
\end{eqnarray}

from the path integral formalism. In this formula, $\hat{\rho}$ is the density matrix of the initial state, $\hat{\Pi} =  \frac{1}{2}[\mathcal{I}-\sqrt{3}\hat{\sigma}_z]$ can be thought of as an inversion operator, and $\hat{U}(\theta,\phi) = e^{-i\frac{\phi}{2}\hat{\sigma}_z} e^{i\frac{\theta}{2}\hat{\sigma}_y}e^{i\frac{\phi}{2}\hat{\sigma}_z}$ is the $SU(2)$ rotation operator.

We consider the time-dependent expectation value of an operator, 

\begin{eqnarray}
\langle\hat{O}(t)\rangle = \langle\psi|\exp\left[i \mathcal{H} t\right]\hat O\exp\left[-i\mathcal{H} t\right]|\psi\rangle
\end{eqnarray}

In the conventional way of constructing the path integral, the time-ordered exponential is split into infinitesimal time steps:

\begin{eqnarray}
\exp\left[i\mathcal{H} t\right] = \lim_{N\rightarrow \infty}\prod_{n=0}^N[1+i\delta\tau\mathcal{H}],
\end{eqnarray}

with $\delta\tau = t/N$. We now insert $2\times(N+1)$ resolutions of the identity

\begin{eqnarray}
\mathcal{I} = \frac{1}{2\pi}\int_0^\pi\sin\theta d\theta \int_0^{2\pi}d\phi |\theta,\phi\rangle\langle\theta,\phi |\equiv \int d\Omega|\Omega\rangle\langle\Omega |,
\end{eqnarray}

with $|\theta,\phi\rangle = \hat{U}(\theta,\phi)|\uparrow\rangle$ the spin coherent state, between each such infinitesimal evolution; this results in

\begin{eqnarray}
\langle\hat{O}(t)\rangle  &=&\lim_{N\rightarrow\infty}\prod_{n=0}^N\int d\Omega^f_n \prod_{n=0}^N\int d\Omega^b_n\langle \psi|\Omega^f_{0}\rangle\times\prod_{n=0}^{N-1}\langle\Omega^f_n|1+i\delta\tau\mathcal{H}|\Omega^f_{n+1}\rangle\nonumber\\
&\times&\langle \Omega^f_{N}|\hat O|\Omega^b_{N}\rangle\nonumber\\
&\times&\prod_{n=0}^{N-1}\langle\Omega^b_{N-n}|1-i\delta\tau\mathcal{H}|\Omega^b_{N-n-1}\rangle\times\langle\Omega^b_0|\psi\rangle.
\end{eqnarray}

We have labeled the introduced identities according to whether they are adjacent to forward or backwards propagation in time. It is well known that the terms proportional to an infinitesimal action of the Hamiltonian produce the path integral:

\begin{eqnarray}
\langle\hat{O}(t)\rangle  &=&\int d\Omega^f_0 d\Omega^f_N \int d\Omega^b_0d\Omega^b_N\langle\Omega^b_0|\psi\rangle\langle \psi|\Omega^f_{0}\rangle\langle\Omega^f_{N}|\hat O|\Omega^b_{N}\rangle\\
&\times& \int D\Omega^f(\tau) \int D\Omega^b(\tau)\exp\left[i\int_0^t d\tau\left(\mathcal{L}(\Omega^f(\tau))-\mathcal{L}(\Omega^b(\tau))\right)\right].\nonumber
\end{eqnarray}

where the Lagrangian density $\mathcal{L}(\Omega(\tau)) = -i\langle\Omega(\tau)|\partial_\tau|\Omega(\tau)\rangle + \langle\Omega(\tau)|\mathcal{H}|\Omega(\tau)\rangle$ is the continuum limit of the infinitesimal time evolution. Note that the integration over the boundary fields (steps $0$ and $N$) has been left in its discrete form; integration over these fields results in the Wigner function for the initial state and the Weyl symbol for the operator $\hat{O}$.

Instead of the fields $\Omega^f$ and $\Omega^b$ propagating forward and backward in time, it is convenient to introduce their classical ($c$) and quantum ($q$) combinations; this is done using the identity

\begin{eqnarray}
\hat{1}\otimes\hat{1} &=& \int d\Omega^f d\Omega^f_b |\Omega^f\rangle\langle\Omega^f|\otimes|\Omega^b\rangle\langle\Omega^b|\nonumber\\
&=&\int d\Omega^c d\Omega^q \hat{U}(\Omega^c)|\Omega^q\rangle\langle\Omega^q|\hat{U}(\Omega^c)^\dagger\otimes\hat{U}(\Omega^c)\hat{\Pi}|\Omega^q\rangle\langle\Omega^q|\hat{\Pi}\hat{U}(\Omega^c)^\dagger,
\end{eqnarray}

which suggests the definitions

\begin{eqnarray}
|\Omega^f\rangle  &=& \hat{U}(\Omega^c)|\Omega^q\rangle,\nonumber\\
|\Omega^b\rangle  &=&\hat{U}(\Omega^c)\hat{\Pi}|\Omega^q\rangle.
\end{eqnarray}

The terms quantum and classical refer to the fact that in the classical evolution all trajectories are uniquely defined and the backward path should be exactly identical to the forward one. This is the case (up to a normalization constant) for $|\Omega^q\rangle = |\uparrow\rangle$ or $|\downarrow\rangle$; for any other value of the quantum field, the two trajectories are different, a manifestation of quantum fluctuations.

We perform this change of variables, and expand the path integral to lowest order in the quantum to field, in order to obtain the truncated Wigner approximation (TWA):

\begin{eqnarray}
\langle\hat{O}(t)\rangle  &\approx&\int d\Omega^c_0 d\Omega^c_N \int d\Omega^q_0d\Omega^q_N\\
&\times&
\langle\Omega^q_0|\hat{\Pi}\hat{U}(\Omega^c_0)^\dagger|\psi\rangle\langle \psi|\hat{U}(\Omega^c_0)|\Omega^q_{0}\rangle\nonumber\\
&\times&\langle\Omega^q_{N}|\hat{U}(\Omega^c_N)^\dagger\hat O\hat{U}(\Omega^c_N)\hat{\Pi}|\Omega^q_{N}\rangle\nonumber\\
&\times& \int D\Omega^c(\tau) \int D\Omega^q(\tau)\exp\left[i\int_0^t d\tau\Omega^q(\tau)\frac{\partial}{\partial\Omega^c(\tau)}\mathcal{L}(\Omega^c(\tau))\right].\nonumber
\end{eqnarray}

Integrating over the quantum variables produces the known components of the TWA:

\begin{eqnarray}
\langle\hat{O}(t)\rangle &\approx&\int d\Omega^c_0  d\Omega^c_N W(\Omega^c_0)O_W(\Omega^c_N)\delta[\Omega^c_N - \Omega_{cl}[\Omega^c_0,t]],
\end{eqnarray}

where

\begin{eqnarray}
W(\Omega^c_0) &=& \int d\Omega^q_0\langle\Omega^q_0|\hat{\Pi}\hat{U}(\Omega^c_0)^\dagger|\psi\rangle\langle \psi|\hat{U}(\Omega^c_0)|\Omega^q_{0}\rangle\nonumber\\
& =& Tr\left[\hat{\rho}\hat{U}(\Omega^c_0)\hat{\Pi}\hat{U}(\Omega^c_0)^\dagger\right]
\end{eqnarray}
is the Wigner function,

\begin{eqnarray}
O_W(\Omega^c_N) &=& \int d\Omega^q_N\langle\Omega^q_{N}|\hat{U}(\Omega^c_N)^\dagger\hat O\hat{U}(\Omega^c_N)\hat{\Pi}|\Omega^q_{N}\rangle\nonumber\\
&=&Tr\left[\hat{O}\hat{U}(\Omega^c_N)\hat{\Pi}\hat{U}(\Omega^c_N)^\dagger\right]
\end{eqnarray}

is the Weyl symbol for the operator $\hat{O}$, and the $\delta$-function over the classical, saddlepoint trajectory is obtained by

\begin{eqnarray}
 \int D\Omega^q(\tau)\exp\left[i\int_0^t d\tau\Omega^q(\tau)\frac{\partial}{\partial\Omega^c(\tau)}\mathcal{L}(\Omega^c(\tau))\right].
\end{eqnarray}

\subsection{Wigner function for an MPS}

Following Ho et al., we start with a different MPS representation of the many-body state, which is also subject to the constraint the no two neighbors are in the excited Rydberg state: $|\psi'({\vartheta,\varphi})\rangle = Tr\left(A'_1A'_2...A'_L\right)$, with

\begin{eqnarray}
A'_i(\vartheta_i,\varphi_i) = \left( {\begin{array}{cc}
   \cos(\vartheta_i/2)|g\rangle & ie^{i\varphi_i}\sin(\vartheta_i/2)|r\rangle \\
    \cos(\vartheta_i/2)|g\rangle & 0 \\
  \end{array} } \right).
\end{eqnarray}

In order to obtain the Wigner function appropriate to our MPS, we consider the identity operator within the restricted Hilbert space:

\begin{eqnarray}
\tilde{1} &=& \prod_{i=1}^L \int_0^\pi d\vartheta_i\sin\vartheta_i\int_0^{2\pi}\frac{d\varphi_i}{2\pi}|\psi'(\vartheta,\varphi)\rangle\langle\psi'(\vartheta,\varphi)|\nonumber\\
 &\equiv& \prod_{i=1}^L \int d \Omega'_i |\Omega'\rangle\langle\Omega'|
\end{eqnarray}

Following the steps of the previous section in obtaining the path integral representation of a correlation function, we find that for our many-body MPS

\begin{eqnarray}
\langle\hat{O}(t)\rangle  &=&\int d\Omega'^f_0 d\Omega'^f_N \int d\Omega'^b_0d\Omega'^b_N\langle\Omega'^b_0|\psi\rangle\langle \psi|\Omega'^f_{0}\rangle\langle\Omega'^f_{N}|\hat O|\Omega'^b_{N}\rangle\\
&\times& \int D\Omega'^f(\tau) \int D\Omega'^b(\tau)\exp\left[i\int_0^t d\tau\left(\mathcal{L}(\Omega'^f(\tau))-\mathcal{L}(\Omega'^b(\tau))\right)\right].\nonumber
\end{eqnarray}

We change variables from forward and backwards to quantum and classical fields; described by these new variables, the matrices constituting the MPS are

\begin{eqnarray}
&&A'(\Omega'^f)\rightarrow \tilde{A}'(\Omega'^c,\Omega'^q) = \\
&&\left( {\begin{array}{cc}
   \left[\cos(\vartheta^c/2)\cos(\vartheta^q/2) + ie^{i(\varphi^c-\varphi^q)}\sin(\vartheta^c/2)\sin(\vartheta^q/2)\right]|g\rangle & \left[-\cos(\vartheta^c/2)\sin(\vartheta^q/2)e^{i\varphi^q} + ie^{i\varphi^c}\sin(\vartheta^c/2)\cos(\vartheta^q/2)\right]|r\rangle \\
    \left[\cos(\vartheta^c/2)\cos(\vartheta^q/2) + ie^{i(\varphi^c-\varphi^q)}\sin(\vartheta^c/2)\sin(\vartheta^q/2)\right]|g\rangle & 0 \\
  \end{array} } \right)\nonumber\\
  &&A'(\Omega'^b)\rightarrow \overline{A}'(\Omega'^c,\Omega'^q) = \frac{1}{2}\times\nonumber\\
&&\left( {\begin{array}{cc}
   \left[(1-\sqrt{3})\cos(\vartheta^c/2)\cos(\vartheta^q/2) + (1+\sqrt{3})ie^{i(\varphi^c-\varphi^q)}\sin(\vartheta^c/2)\sin(\vartheta^q/2)\right]|g\rangle & \left[-(1-\sqrt{3})\cos(\vartheta^c/2)\sin(\vartheta^q/2)e^{i\varphi^q}\right.\nonumber\\
    &\left. + (1+\sqrt{3})ie^{i\varphi^c}\sin(\vartheta^c/2)\cos(\vartheta^q/2)\right]|r\rangle \\
    \left[(1-\sqrt{3})\cos(\vartheta^c/2)\cos(\vartheta^q/2) +(1+\sqrt{3}) ie^{i(\varphi^c-\varphi^q)}\sin(\vartheta^c/2)\sin(\vartheta^q/2)\right]|g\rangle & 0 \\
  \end{array} } \right)\nonumber
\end{eqnarray}

where, if we define

\begin{eqnarray}
|\psi_1(\Omega'^c,\Omega'^q)\rangle &=& Tr(\tilde{A}'_1\tilde{A}'_2...\tilde{A}'_L)\nonumber\\
|\psi_2(\Omega'^c,\Omega'^q)\rangle &=& Tr(\overline{A}'_1\overline{A}'_2...\overline{A}'_L)
\end{eqnarray}

it can be shown that

\begin{eqnarray}
\tilde{1}\otimes\tilde{1} &=& \int d\Omega'^f d\Omega'^f_b |\Omega'^f\rangle\langle\Omega'^f|\otimes|\Omega'^b\rangle\langle\Omega'^b|\\
&=&\int d\Omega'^c d\Omega'^q |\psi_1(\Omega'^c,\Omega'^q)\rangle\langle\psi_1(\Omega'^c,\Omega'^q)|\otimes|\psi_2(\Omega'^c,\Omega'^q)\rangle\langle\psi_2(\Omega'^c,\Omega'^q)|\nonumber.
\end{eqnarray}

This leads us, in the spirit of the previous section, to a definition of the Wigner function for the MPS:

\begin{eqnarray}
\tilde{W}(\Omega'^c_0) = \int d\Omega'^q_0\langle\psi_2(\Omega_0'^c,\Omega_0'^q)|\psi\rangle\langle \psi|\psi_1(\Omega_0'^c,\Omega_0'^q)\rangle.
\end{eqnarray}

In particular, for the $|Z_2\rangle = |g_1\rangle|r_2\rangle$ initial state, where we are interested in only two sets of variables $\vartheta_{1,2}, \varphi_{1,2}$, the Wigner function is given by

\begin{eqnarray}
\tilde{W}(\vartheta_{1,2},\varphi_{1,2}) &=& \frac{1}{(2\pi)^2}\int_0^\pi d\vartheta_1^q\sin(\vartheta_1^q)\int_0^\pi d\vartheta_2^q\sin(\vartheta_2^q)\int_0^{2\pi}d\varphi_1^q\int_0^{2\pi}d\varphi_2^q\\
&\times& \left[\cos(\vartheta_1/2)\cos(\vartheta_1^q/2) + ie^{i(\varphi_1-\varphi_1^q)}\sin(\vartheta_1/2)\sin(\vartheta_1^q/2)\right]\nonumber\\
&\times&\left[(1-\sqrt{3})\cos(\vartheta_1/2)\cos(\vartheta_1^q/2) - (1+\sqrt{3})ie^{-i(\varphi_1-\varphi_1^q)}\sin(\vartheta_1/2)\sin(\vartheta_1^q/2)\right]\nonumber\\
&\times&\left[-\cos(\vartheta_2/2)\sin(\vartheta_2^q/2)e^{i\varphi_2^q} + ie^{i\varphi_2}\sin(\vartheta_2/2)\cos(\vartheta_2^q/2)\right]\nonumber\\
&\times&\left[-(1-\sqrt{3})\cos(\vartheta_2/2)\sin(\vartheta_2^q/2)e^{-i\varphi_2^q} - (1+\sqrt{3})ie^{-i\varphi_2}\sin(\vartheta_2/2)\cos(\vartheta_2^q/2)\right]\nonumber\\
&=&\frac{1}{4}[1-\sqrt{3}\cos(\vartheta_1)][1+\sqrt{3}\cos(\vartheta_2)]
\end{eqnarray}

\subsection{Transformation to new variables}

We now perform a transformation to different variables $\theta,\phi$, which correspond to the MPS $|\psi(\theta,\phi)\rangle = Tr\left(A_1A_2...A_L\right)$ with

\begin{eqnarray}
A_i(\theta_i,\phi_i) = \left( {\begin{array}{cc}
   \cos(\theta_i/2)|g\rangle & ie^{i\phi_i}\sin(\theta_i/2)|r\rangle \\
    |g\rangle & 0 \\
  \end{array} } \right).
\end{eqnarray}

This representation is favorable as the state is normalized, and the equations of motion which give rise to the quantum scar are known in terms of $\theta,\phi$ rather than $\vartheta,\varphi$. The difference between the two representations is that the amplitude for a ground state atom to occure on site $i+1$, given that an excited Rydberg atom resides on site $i$, is given by $\cos(\vartheta_{i+1}/2)$ for $|\psi'(\vartheta,\varphi)\rangle$, and by $1$ for $|\psi(\theta,\phi)\rangle$. The transition between the two representations is achieved via a gauge transformation, as detailed in Ho et al.

In the subset of states which are invariant under a translation by two sites, i.e. $\vartheta_{i+2} = \vartheta_i$ et cetera, the gauge transformation becomes

\begin{eqnarray}
A'(\vartheta,\varphi)\rightarrow \frac{\cos(\vartheta/2)}{\cos(\theta/2)}A(\theta,\phi),
\end{eqnarray}
where the variables are related by
\begin{eqnarray}
\cos^2(\theta_1/2) &=& 2\left[1+\tan^2(\vartheta_1/2)-\tan^2(\vartheta_2/2)+\sqrt{4\tan^2(\vartheta_2/2)+(1+\tan^2(\vartheta_1/2)-\tan^2(\vartheta_2/2))^2}\right]^{-1}\nonumber\\
\tan(\vartheta_1/2) &=& \tan(\theta_1/2)/\cos(\theta_2/2).
\end{eqnarray}

The Jacobian for this transformation is
\begin{eqnarray}
J(\theta_1,\theta_2) = \frac{1 - \sin^2\left(\frac{\theta_1}{2}\right)\sin^2\left(\frac{\theta_2}{2}\right)}{\left[\cos^2\left(\frac{\theta_1}{2}\right) + \tan^2\left(\frac{\theta_2}{2}\right)\right]\left[\cos^2\left(\frac{\theta_2}{2}\right) + \tan^2\left(\frac{\theta_1}{2}\right)\right]\cos\left(\frac{\theta_1}{2}\right)\cos\left(\frac{\theta_2}{2}\right)}
\end{eqnarray}

This leads us to a representation of the density matrix via a Wigner function of the new variables:

\begin{eqnarray}
W(\theta_1,\theta_2) = J(\theta_1,\theta_2)\frac{\sin\left(\vartheta_1(\theta_1,\theta_2)\right)\sin\left(\vartheta_2(\theta_1,\theta_2)\right)}{\sin(\theta_1)\sin(\theta_2)}\tilde{W}(\vartheta_1(\theta_1,\theta_2),\vartheta_2(\theta_1,\theta_2)),
\end{eqnarray}

where the Jacobian and the sine functions have been introduced so that the Wigner function obeys the Stratonovich - Weyl condition

\begin{eqnarray}
\int_0^\pi d\theta_1\sin(\theta_1)\int_0^\pi d\theta_2\sin(\theta_2)W(\theta_1,\theta_2) &=& \int_0^\pi d\vartheta_1\sin(\vartheta_1)\int_0^\pi d\vartheta_2\sin(\vartheta_2)\tilde{W}(\vartheta_1,\vartheta_2) \nonumber\\
&=&Tr[\rho] = 1.
\end{eqnarray}

This leads us to the expression given in the paper.

\section{Lyapunov exponents}

\subsection{Equations of motion for unit cell of size $L$}

We obtain the Lagrangian for a system with periodicity, or unit cell size, $L$, where $L$ is even; the $L=2$ case has been studied previously, and the resulting equations of motion for the corresponding variables $\theta_2,\theta_1$ are shown in the paper.

The state of the system, in our bond dimension $\chi = 2$ approximation, is given by

\begin{eqnarray}
|\psi[\{\theta_i,\phi_i\}]\rangle = \sum_{\{\sigma_i\}}v_L^\dagger A(\theta_1,\phi_1)^{\sigma_1}A(\theta_2,\phi_2)^{\sigma_2}...A(\theta_N,\phi_N)^{\sigma_N}v_R|\{\sigma_i\}\rangle,
\end{eqnarray}

with $\theta_{i+L} = \theta_i$, $\phi_{i+L} = \phi_i$, and the matrices $A(\theta,\phi)$ given in the paper. We consider an infinite system, meaning the limit $N\rightarrow \infty$, and calculate quantities to leading order in $1/N$.

The Lagrangian within the TDVP, for a system of such periodicity, is given by

\begin{eqnarray}
\mathcal{L}[\{\theta_i(t),\phi_i(t)\}] &=& i\sum_{m=1}^L\dot{\theta}_m\langle \psi[\{\theta_i,\phi_i\}]|\frac{\partial}{\partial\theta_m}|\psi[\{\theta_i,\phi_i\}]\rangle\nonumber\\
&+& i\sum_{m=1}^L\dot{\phi}_m\langle \psi[\{\theta_i,\phi_i\}]|\frac{\partial}{\partial\phi_m}|\psi[\{\theta_i,\phi_i\}]\rangle\nonumber\\
&-&\langle \psi[\{\theta_i,\phi_i\}]|H|\psi[\{\theta_i,\phi_i\}]\rangle.
\end{eqnarray}

In our calculation of quantities such as $\langle \psi|\frac{\partial}{\partial\theta_m}|\psi\rangle$, we encounter expressions of the form

\begin{eqnarray}
&&\Upsilon \equiv \\
&&\sum_{\{\sigma_i\}}v_R^\dagger A(\theta_N,\phi_N)^{\sigma_N\dagger}...A(\theta_1,\phi_1)^{\sigma_1\dagger}v_Lv_L^\dagger A(\theta_1,\phi_1)^{\sigma_1}...\partial_{\theta_m}A(\theta_m,\phi_m)^{\sigma_m}...A(\theta_N,\phi_N)^{\sigma_N}v_R;\nonumber
\end{eqnarray}

we therefore need to understand how to solve for matrices such as

\begin{eqnarray}
\begin{pmatrix}
    P       & 0 \\
    0       & Q \\
\end{pmatrix} = \sum_{\{\sigma_i\}}A(\theta_m,\phi_m)^{\sigma_m\dagger} ...A(\theta_1,\phi_1)^{\sigma_1\dagger}v_Lv_L^\dagger A(\theta_1,\phi_1)^{\sigma_1}...\partial_{\theta_m}A(\theta_m,\phi_m)^{\sigma_m},
\end{eqnarray}

such that 

\begin{eqnarray}
&&\Upsilon =  \\
&&\sum_{\{\sigma_i\}}v_R^\dagger A(\theta_N,\phi_N)^{\sigma_N\dagger}...A(\theta_{m+1},\phi_{m+1})^{\sigma_1\dagger}
\begin{pmatrix}
    P       & 0 \\
    0       & Q \\
\end{pmatrix}
A(\theta_{m+1},\phi_{m+1})^{\sigma_{m+1}}...A(\theta_N,\phi_N)^{\sigma_N}v_R\nonumber
\end{eqnarray}

The basic identity we use is
\begin{eqnarray}\label{eq:basic}
\sum_{\sigma}A^{\sigma\dagger}(\theta)
\begin{pmatrix}
    \alpha       & 0 \\
    0       & \beta \\
\end{pmatrix}A^{\sigma}(\theta) = 
\begin{pmatrix}
    \alpha\cos^2\theta+\beta       & 0 \\
    0       &\alpha\sin^2\theta \\
\end{pmatrix};
\end{eqnarray}
note that the trace of the matrix is preserved. We therefore find that

\begin{eqnarray}
&&\Upsilon = 
v_R^\dagger 
\begin{pmatrix}
    P       & 0 \\
    0       & Q \\
\end{pmatrix}
v_R = P+Q.
\end{eqnarray}

We now turn to a calculation of the parameters $P$ and $Q$, by considering 

In a size-$L$ unit cell we get, after a series of the $L$ different matrices
\begin{eqnarray}
\sum_{{\sigma_1,\sigma_2...\sigma_L}}A^{\sigma_L\dagger}(\theta_L)...A^{\sigma_1\dagger}(\theta_1)
\begin{pmatrix}
    \alpha       & 0 \\
    0       & 1-\alpha \\
\end{pmatrix}A^{\sigma_1}(\theta_1)...A^{\sigma_L}(\theta_L) = 
\begin{pmatrix}
   A       & 0 \\
    0       &1-A 
\end{pmatrix},
\end{eqnarray}
with $A = 1 - \sin^2\theta_L + \sin^2\theta_L\sin^2\theta_{L-1}-...-\alpha\prod_{i=1...L}\sin^2\theta_i$. This leads us to define, in a recursive way, 

We define, in a recursive way, the 
\begin{eqnarray}\label{eq:recursion1}
\begin{pmatrix}
    \alpha_{m,n+1}       & 0 \\
    0       &1-\alpha_{m,n+1}  \\
\end{pmatrix}
&=&
\sum_{\sigma_1,\sigma_2...\sigma_L} A^{\sigma_{m-1}\dagger}(\theta_{m-1})...A^{\sigma_{1}\dagger}(\theta_1)A^{\sigma_L\dagger}(\theta_L)...A^{\sigma_{m+1}\dagger}(\theta_{m+1})A^{\sigma_m\dagger}(\theta_m)\nonumber\\
&\times&
\begin{pmatrix}
    \alpha_{m,n}       & 0 \\
    0       &1-\alpha_{m,n} \\
\end{pmatrix}\nonumber\\
&\times& A^{\sigma_m}(\theta_m)A^{\sigma_{m+1}}(\theta_{m+1})...A^{\sigma_L}(\theta_L)A^{\sigma_{1}}(\theta_1)...A^{\sigma_{m-1}}(\theta_{m-1}).
\end{eqnarray}

A simple calculation shows that
\begin{eqnarray}\label{eq:recursion2}
\alpha_{m,n} = \frac{1-\Pi^n}{1-\Pi}a_m + \alpha_{m,1}\Pi^{n-1},
\end{eqnarray}
with $\Pi = \prod_i\sin^2\theta_i$, and $a_m = 1+\sum_{k=1}^{L-1} (-1)^{k} \prod_{l=1}^k \sin^2\theta_{m-l}$, where the subscript of $\theta$ is defined $\mod(L)$.

This result allows us to calculate derivative terms such as $\langle \psi|\frac{\partial}{\partial\theta_m}|\psi\rangle$; the derivative with respect to $\theta_m$ acts on the matrices $A(\theta_m,\phi_m)$ which appear with periodicity $L$ in the definition of the state $|\psi\rangle$. The derivative acting on the first such matrix results in (using the basic identity Eq.\ref{eq:basic})

\begin{eqnarray}
v_R^\dagger \sum_{\sigma_m}A^{\sigma_m\dagger}
\begin{pmatrix}
    \beta       & 0 \\
    0       &1-\beta \\
\end{pmatrix}
\partial_{\theta_m}A^{\sigma_m}v_R = \beta\cos(\theta_m)\sin(\theta_m) \times v_R^\dagger
\begin{pmatrix}
   -1       & 0 \\
    0       &1 \\
\end{pmatrix}
v_R =  0,
\end{eqnarray}

while the derivative with resepct to $\phi_m$ on such a matrix is given by

\begin{eqnarray}
v_R^\dagger \sum_{\sigma_m}A^{\sigma_m\dagger}
\begin{pmatrix}
    \beta       & 0 \\
    0       &1-\beta \\
\end{pmatrix}
\partial_{\phi_m}A^{\sigma_m}v_R =  iv_R^\dagger
\begin{pmatrix}
   0        &0 \\
    0       &\beta\sin^2(\theta_m) \\
\end{pmatrix}
v_R =  \beta\sin^2(\theta_m),
\end{eqnarray}

where $\beta$ is the upper diagonal of the matirx $\sum_{\sigma_1...\sigma_{m-1}} A^{\sigma_{m-1}\dagger}...A^{\sigma_{1}\dagger} v_Lv_L^\dagger A^{\sigma_{1}}...A^{\sigma_{m-1}}$.

Using Eqs.\ref{eq:recursion1},\ref{eq:recursion2}, we are able to calculate the contribution from the derivative of the matrix in the $n$-th unit cell; summing up all these contributions results in

\begin{eqnarray}
\langle \psi|\frac{\partial}{\partial\theta_m}|\psi\rangle &=& \sin^2(\theta_m) \sum_n
\left[\frac{1-\Pi^n}{1-\Pi}a_m + \beta \Pi^{n-1}\right]\nonumber\\
&=&N\sin^2(\theta_m)\frac{a_m}{1-\Pi} + O(1).
\end{eqnarray}

Calculating the Hamiltonian term follows the same line of reasoning.

These considerations result in the following Lagrangian, to leading order in the system size:
\begin{eqnarray}
-\frac{1}{N} \mathcal{L} = \sum_m A_m\sin^2\theta_m\dot\phi_m + \sum_mA_m\sin\theta_m\cos\theta_m\cos\theta_{m+1}\sin\phi_m
\end{eqnarray}

where $A_m = a_m/(1-\Pi)$.

This Lagrangian results in the following equation of motion:
\begin{eqnarray}
\dot\theta_i = M^{-1}_{ij}u_j,
\end{eqnarray}
with $M_{ij} = \frac{\partial \Phi_i}{\partial \theta_j}$ and $u_i = -A_m\sin\theta_i\cos\theta_i\cos\theta_{i+1}$, with $\Phi_i = A_i\sin^2\theta_i$. The inverse of $M$ is given by $M^{-1}_{ij} = \frac{\partial \theta_i}{\partial \Phi_j}$. Using the identity $A_m\sin^2\theta_m =1 - A_{m+1}$, we get $\theta_i = \sin^{-1}\left(\sqrt{\frac{\Phi_k}{1-\Phi_{k-1}}}\right)$, and thus we end with the equations of motion $\dot\theta_i = M^{-1}_{ij}u_j,$ with
\begin{eqnarray}
M^{-1}_{ij} = \delta_{i,j}\frac{1}{2A_i\sin\theta_i\cos\theta_i} + \delta_{i-1,j}\frac{1}{2}\frac{1}{A_i}\frac{\sin\theta_i}{\cos\theta_i}
\end{eqnarray}

which gives
\begin{eqnarray}
\dot\theta_i = -\frac{1}{2}\left[\cos\theta_{i+1} + \frac{A_{i-1}}{A_i}\sin\theta_{i-1}\cos\theta_{i-1}\sin\theta_i\right]\equiv f_i(\{\theta\}).
\end{eqnarray}

\subsection{Calculation of Lyapunov exponents}
The equations of motion for small perturbations are given by
\begin{eqnarray}
\delta\dot\theta_i = \frac{\partial f_i}{\partial \theta_j}\delta\theta_j,
\end{eqnarray}
and thus, after one period of the oscillation, the deviation is
\begin{eqnarray}
\delta\theta_i(t) = \mathcal{T}\exp\left[\int_0^\tau \frac{\partial f_i(\{\theta(t)\})}{\partial \theta_j}d\tau\right]\delta\theta_j(t=0)\equiv T_{ij}\delta\theta_j(t=0) ,
\end{eqnarray}
where $\mathcal{T}$ stands for time ordering.

For a given unit cell size $L$, we calculate $\tilde F_{ij} = \frac{\partial f_i}{\partial \theta_j}$. We then revert to the $Z_2$ invariant representation by setting all odd $\theta$'s to $\theta_1$, and all even $\theta$'s to $\theta_2$: $F_{ij} = \tilde F_{ij}|_{\theta_{i}\rightarrow \theta_{\mod(i,2)}}.$ 

Symmetries of the $F$ matrix:
\begin{enumerate}
\item $F_{i+2,j+2} = F_{i,j}$.
\item $F(\theta_2,\theta_1)_{i+1,j+1} = F(\theta_1,\theta_2)$
\end{enumerate}

For simplicity, we use only the lowest harmonic in the numerical solution of the $Z_2$ equations of motion, which provide a very good approximation:
\begin{eqnarray}
\theta_1(t) &=& -\frac{\pi}{2}\left(1-\cos \omega t\right)\nonumber\\
\theta_2 (t) &=& \frac{\pi}{2}\left(1-\sin \omega t\right),
\end{eqnarray}
where $\omega$ is the numerically obtained frequency of the orbit. To confirm that this approximaiton gives reasonable results, we claculate the Lyapunov exponents brute-force by comparing two nearby trajectories and measuring their divergence; such a scheme results in Lyapunov exponents of the same order of magnitude.

We then calculate the Lyapunov exponent in the following manner. We divide the time into sections of size $dt$. We compute the time-ordered exponential
\begin{eqnarray}
Q_{1/8} =\exp[F(\{\theta(t=Ndt)\})]...\times\exp[F(\{\theta(t=3dt)\})]\times\exp[F(\{\theta(t=2dt)\})]\times\exp[F(\{\theta(t=dt)\})],
\end{eqnarray}
such that $Ndt = \tau/8$, where $\tau$ is the period. 
We then use the symmetries of the first harmonic approximation, namely
\begin{eqnarray}
\theta_1(1/4\tau-t) &=& -\theta_2(t); \theta_2(1/4\tau-t) = -\theta_1(t)\nonumber\\
\theta_1(t+1/4\tau) &=& \theta_2(t)-\pi; \theta_2(t+1/4\tau) = -\theta_1(t)\nonumber\\
\theta_1(t+1/2\tau) &=&\pi-\theta_1(t); \theta_2(t+1/2\tau) = \pi-\theta_2(t)
\end{eqnarray}

To conclude that
\begin{eqnarray}
T_{ij} =[S_y\times S_x \times Q_{1/8}\times S_x^{-1}\times  Q_{1/8}\times S_y^{-1}\times S_x\times Q_{1/8}\times S_x^{-1}\times Q_{1/8}]^2,
\end{eqnarray}

where $S_{x,mn} = \delta_{n,m+1}, S_{y,mn} = (-1)^mi\delta_{n,m+1}$.

The Lyapunov exponents are defined as $\lambda = \frac{1}{\tau}\log[\mbox{eig}(T)]$, where eig$(T)$ are the eigenvalues of the matrix $T_{ij}$.

\end{document}